\documentclass[prl,twocolumn,superscriptaddress,noshowpacs]{revtex4}
\usepackage{graphicx}
\usepackage{physics}
\usepackage{bm}
\usepackage{ulem}
\usepackage{color}

\begin{document}

\title{Topological moiré polaritons}

\author{I.~Septembre}
\affiliation{Institut Pascal, PHOTON-N2, Universit\'e Clermont Auvergne, CNRS, Clermont INP, F-63000 Clermont-Ferrand, France}
\author{C.~Leblanc}
\affiliation{Institut Pascal, PHOTON-N2, Universit\'e Clermont Auvergne, CNRS, Clermont INP, F-63000 Clermont-Ferrand, France}
\affiliation{Minatec Campus, Leti, CEA, Grenoble, 38054, France}
\author{D.~D.~Solnyshkov}
\affiliation{Institut Pascal, PHOTON-N2, Universit\'e Clermont Auvergne, CNRS, Clermont INP, F-63000 Clermont-Ferrand, France}
\affiliation{Institut Universitaire de France (IUF), F-75231 Paris, France}
\author{G.~Malpuech}
\affiliation{Institut Pascal, PHOTON-N2, Universit\'e Clermont Auvergne, CNRS, Clermont INP, F-63000 Clermont-Ferrand, France}

\begin{abstract}
The combination of an in-plane honeycomb potential and of a photonic spin-orbit coupling (SOC) emulates a photonic/polaritonic analog of bilayer graphene. We show that modulating the SOC magnitude allows to change the overall lattice periodicity, emulating any type of moiré-arranged bilayer graphene with a unique all-optical access to the moiré band topology. We show that breaking the time-reversal symmetry by an effective exciton-polariton Zeeman splitting opens a large topological gap in the array of moiré flat bands. This gap contains one-way topological edge states whose constant group velocity makes an increasingly sharp contrast with the flattening moiré bands. 
\end{abstract}


\maketitle 
The study of moiré van der Waals bilayers of two-dimensional (2D) materials became a very active field of research in the last decade \cite{andrei2020graphene,he2021moire,huang2022excitons,mak2022semiconductor,du2023moire}. It allows inducing a large-scale periodicity controlled by the moiré angle and described by the moiré factor. These 2D superlattices demonstrate a reduced Brillouin zone with a large number of folded bands. The interaction between these folded bands makes them flatter, whereas the non-trivial geometry of the original van der Waals bilayer bands makes some of these bands topologically non-trivial \cite{Liu2019,wu2019topological,li2021quantum}. From the experimental point of view, the small energy size of the bands allowed to control the superlattice band filling easily and to demonstrate behaviour typical for strongly-correlated electronic phases \cite{Hererro2018}, induced by the unique properties of topological flat bands \cite{Liu2012,Yang2012,bergholtz2013topological,Parameswaran2013,wu2021chern}. 

On the other hand, emulating the Hamiltonians of solid state physics using alternative platforms (atomic, photonic, phononic, electronic, mechanical, etc.) also became a popular research field \cite{Bissbort2013,gross2017quantum,he2016acoustic,Ozawa2019,xue2022topological,wang2020circuit}. In particular, direct emulation of moiré superlattices has been successfully achieved in photonics \cite{mao2021magic,wang2020localization,wang2022two}. Synthetic topological matter allowed for achieving effects and phases challenging to implement in electronic systems. It can allow to reach inaccessible parameters and designs, or to perform measurements out of reach in solid state physics systems, such as directly measuring a wave function, or the geometry of a topological band. An interesting platform from this point of view is the one based on coupled photonic micropillars, which host 0D atomic-like photonic modes \cite{Bajoni2008}. Different types of 1D and 2D lattices (Lieb, honeycomb, etc) have been implemented by making periodic arrays of these pillars \cite{Jacqmin2014,Whittaker2018,klembt2018exciton}. One advantage of this specific platform is that it allows for direct measurement of the band dispersion, the wave functions, and possibly the full band geometry, even if the latter has only been measured in planar microcavities, so far \cite{gianfrate2020measurement}. This platform also allows to strongly couple photonic modes to excitons forming cavity exciton-polaritons (polaritons) \cite{Microcavities}. This strongly-coupled system demonstrates a polaritonic gain and was the ground for the first proposal \cite{Solnyshkov2016} and demonstration \cite{StJean2017} of a topological laser in the edge mode of a polaritonic Su-Schrieffer-Heeger chain \cite{Su1979}. The interacting character of polaritons allows us to address the physics of interacting quantum fluids \cite{carusotto2013quantum} in topologically non-trivial systems \cite{solnyshkov2021microcavity}. The recent observations of strongly interacting regimes for various cavity polariton systems \cite{munoz2019emergence,zhang2021van,zasedatelev2021single} are promising for the achievement of strongly correlated phases and quantum computing. 

Two distinct types of spin-orbit coupling (SOC) have been identified in microcavities. The first has been the TE-TM SOC \cite{Kavokin2005,Tercas2014}, whose effective field scales as $k^2$ where $k$ is the wave vector modulus, and rotates twice faster than $\phi$, the wave vector polar angle. The second is the more recently discovered Rashba-Dresselhaus SOC \cite{rechcinska2019engineering,Ren2021}, which is proportional to the projection of the wave vector on a single axis determining the orientation of emergent optical activity. Interestingly, the effective Hamiltonian describing a honeycomb lattice in the presence of the TE-TM SOC is, close to the Dirac point, very similar to the one of bilayer graphene \cite{Nalitov2015b}, where the two layers are represented by the coupled spin components describing the polarisation of light, which plays the role of a synthetic dimension. Time-reversal symmetry can be broken by the excitonic Zeeman splitting, which opens a topological gap implementing a photonic analog of quantum anomalous Hall effect \cite{Nalitov2015}, recently observed experimentally \cite{klembt2018exciton}. The same combination of exciton SOC (induced by their coupling to the TE and TM photonic modes) and Zeeman splitting has allowed predicting the topological nature of excitons in moiré bilayers of transitional metal dichalcogenides \cite{MacDonald2017}.

In this work, we consider theoretically a spatial modulation of the magnitude of the TE-TM SOC in a polaritonic honeycomb lattice. This induces a large-scale spatial periodicity to the coupling between the effective "layers" (the two spin components), creating an optical analog of a moiré-arranged bilayer graphene. We analyze the band properties in the absence and in the presence of a Zeeman splitting, which opens a topological gap between two groups of flat bands. We show that the band width exhibits an overall decay with the moiré parameter $\alpha$ (inverse of the moiré angle), accompanied by a series of periodic minima at "magic angles". We demonstrate that the flat bands are topological by calculating their Berry curvature distribution, which exhibits symmetry switching related to the periodic minima. Finally, we study the topological edge states and demonstrate their robustness, leading to an exceptionally high contrast between the propagative edge states with high group velocity and the flat bands of the bulk.

The polariton graphene with TE-TM spin-orbit coupling is well described by the following tight-binding Hamiltonian \cite{Nalitov2015,Nalitov2015b,klembt2018exciton}:
\begin{equation}\label{Ham_H}
\mathrm{H}_\mathbf{k} = \left( \begin{matrix}
\Delta \sigma_z & \mathrm{F}_{\mathbf{k}} \\
\mathrm{F}_{\mathbf{k}}^\dagger & \Delta \sigma_z
\end{matrix} \right), \quad
\Delta = \vert x \vert ^2 g_X \mu_B H_z/2,
\end{equation}
This Hamiltonian is written on the $(A^+,A^-,B^+,B^-)$ basis, where $A$ and $B$ are the sublattice sites and $(+,-)$ stands for the circular polarisation (spin) of light.
 $\sigma_z$ is the Pauli matrix, $x$ is the excitonic Hopfield coefficient, $g_X$ is the effective g-factor for the 2D exciton, $\mu_B$ is the Bohr magneton, and $H_z$ is the applied magnetic field, giving rise to polariton Zeeman splitting $\Delta$.
 \begin{equation} \label{Hamiltonian}
\mathrm{F}_{\mathbf{k}} = - \left( \begin{matrix}
f_{\mathbf{k}} J & f_{\mathbf{k}}^+ \delta J \\
f_{\mathbf{k}}^- \delta J & f_{\mathbf{k}} J
\end{matrix} \right),
\end{equation}
where complex coefficients $f_{\mathbf{k}}$,$f_{\mathbf{k}}^\pm$ are defined by:
\begin{equation}
f_{\mathbf{k}}=\sum_{j=1}^3 \exp(-\mathrm{i}\mathbf{k d}_{\varphi_j}),\quad
f_{\mathbf{k}}^\pm = \sum_{j=1}^3 \exp(-\mathrm{i}\left[\mathbf{k d}_{\varphi_j} \mp 2 \varphi_j \right]), \notag
\end{equation}
and $\varphi_j = 2 \pi (j-1) / 3$ is the angle between the horizontal axis and the direction to the $j$th nearest neighbor of a type-A pillar.
$J$ is the polarisation-independent tunneling coefficient, whereas $\delta J$ is the polarisation-dependent term induced by the TE-TM SOC \cite{Sala2015,Nalitov2015}.

The two spin components in this Hamiltonian are analogous to the two layers of the bilayer graphene \cite{thesis2018bleu,gianfrate2020measurement,mccann2013electronic}. Modulating the coupling between them is, therefore, qualitatively similar to the moiré modulation arising in twisted bilayer graphene \cite{dosSantos2007,Moon2012,bistritzer2011moire}. To show this, we focus on the Hamiltonian~\eqref{Ham_H} close to the Dirac point, which in the first order reads
\begin{widetext}
\begin{equation}
\label{TETMnonmod}
H\approx \left( {\begin{array}{*{20}{c}}
0&0&{\frac{{3d}}{2}J\left( {\frac{\partial }{{\partial x}} - i\frac{\partial }{{\partial y}}} \right)}&{\frac{{3d}}{2}\delta J\left( {\frac{\partial }{{\partial x}} + i\frac{\partial }{{\partial y}}} \right)}\\
0&0&{ - 3\delta J}&{\frac{{3d}}{2}J\left( {\frac{\partial }{{\partial x}} - i\frac{\partial }{{\partial y}}} \right)}\\
{\frac{{3d}}{2}J\left( {\frac{\partial }{{\partial x}} + i\frac{\partial }{{\partial y}}} \right)}&{ - 3\delta J}&0&0\\
{\frac{{3d}}{2}\delta J\left( {\frac{\partial }{{\partial x}} - i\frac{\partial }{{\partial y}}} \right)}&{\frac{{3d}}{2}J\left( {\frac{\partial }{{\partial x}} + i\frac{\partial }{{\partial y}}} \right)}&0&0
\end{array}} \right)
\end{equation}
\end{widetext}
where we have replaced the wave vectors with the corresponding operators $k_x=-i\partial/\partial x$. The upper right (and lower left) term of the matrix is the trigonal warping term \cite{Dresselhaus1974,Nalitov2015b}. If it is absent, the dispersion near the Dirac point is composed of two touching parabolas. When this term is present, the parabolas are crossing, giving rise to four Dirac points, which is precisely the effect called trigonal warping. The Hamiltonian~\eqref{TETMnonmod} corresponds exactly to the one of an AB-stacked bilayer graphene \cite{mccann2013electronic} with $\gamma_4=0$ (the term corresponding to tunnelling from a $A$ site in one layer to the $A$ site in the second layer). This correspondence is the starting point of our analogy for moiré emulation.

We now consider the moiré modulation of the TE-TM coupling:
\begin{widetext}
\begin{equation}
H\approx \left( {\begin{array}{*{20}{c}}
0&0&{\frac{{3d}}{2}J\left( {\frac{\partial }{{\partial x}} - i\frac{\partial }{{\partial y}}} \right)}&{\frac{{3d}}{2}\delta JV(x,y)\left( {\frac{\partial }{{\partial x}} + i\frac{\partial }{{\partial y}}} \right)}\\
0&0&{ - 3\delta JV(x,y)}&{\frac{{3d}}{2}J\left( {\frac{\partial }{{\partial x}} - i\frac{\partial }{{\partial y}}} \right)}\\
{\frac{{3d}}{2}J\left( {\frac{\partial }{{\partial x}} + i\frac{\partial }{{\partial y}}} \right)}&{ - 3\delta JV(x,y)}&0&0\\
{\frac{{3d}}{2}\delta JV(x,y)\left( {\frac{\partial }{{\partial x}} - i\frac{\partial }{{\partial y}}} \right)}&{\frac{{3d}}{2}J\left( {\frac{\partial }{{\partial x}} + i\frac{\partial }{{\partial y}}} \right)}&0&0
\end{array}} \right)
\label{effMoireTETM}
\end{equation}
\end{widetext}
where $V(x,y)=(M+(1/3+2/9\cos(\sqrt{3}ak_y/m)+2/9\cos(a(3k_x-\sqrt{3}k_y)/2m)+2/9\cos(a(3k_x+\sqrt{3}k_y)/2m)))/M$ is the modulation term with $m\in \mathcal{Z}$ being the moiré parameter and $1/M$ being the moiré modulation amplitude.
The moiré parameter $m$ is linked with the moiré angle $\theta$ by $2\sin\theta/2=1/m$ \cite{Moon2012}. The modulation could be achieved during the fabrication of the microcavity by adding a spatially-periodic asymmetry to the mirrors without changing the cavity thickness.
This should be compared with the effective Hamiltonian (2) of Ref.~\cite{Tarnopolsky2019}, which has the simplest shape and the most pronounced behavior (“absolutely flat” bands). The difference is that we have modulated spatial derivatives in one of the terms, and that the modulation is real in our case and complex in their case. Moreover, the two terms are comparable in their case, whereas the one with spatial derivatives is actually negligible in our case. So, while qualitatively, the moiré modulation is well present in both cases with the same period, there are also important differences, the consequences of which we will study below.

Next, we go beyond the effective Hamiltonian approximation (whose results are presented in the Supplemental Materials \cite{suppl}) by considering the full tight-binding model of graphene lattice with modulated TE-TM SOC, similar to some works on moiré bilayers \cite{Moon2012}. The Hamiltonian reads:
\begin{eqnarray}
\label{HTBm}
    H_{TBm}&=&-\sum\limits_{l,n} J_{ln}\ket{l,\pm}\bra{n,\pm}\\&-&\sum\limits_{l,n}\delta J_{ln}e^{-2i\varphi_{ln}}\ket{l,\pm}\bra{n,\mp}+H.c.\nonumber
\end{eqnarray}
Here, the atom numbers $l,n$ scan the moiré BZ, the spin-conserving and spin-flipping tunnelings $J_{l,n}$ and $\delta J_{l,n}$ are nonzero only for nearest neighbors, $\varphi_{l,n}$ is the direction of the link $l\to n$, and the TE-TM spin-flipping tunneling is modulated as $\delta J=\delta J_0(M-1+\cos(\pi x_{l,n}/mb_x)\cos(\pi y_{l,n}/mb_y))/M$, where $1/M=5\%$ is the modulation strength, $\bm{b}$ is one of the basis vectors of the basic unit cell. This modulation is a tight-binding version of $V(x,y)$ introduced above.

\begin{figure}[tbp]
\centering
\includegraphics[width=0.99\linewidth]{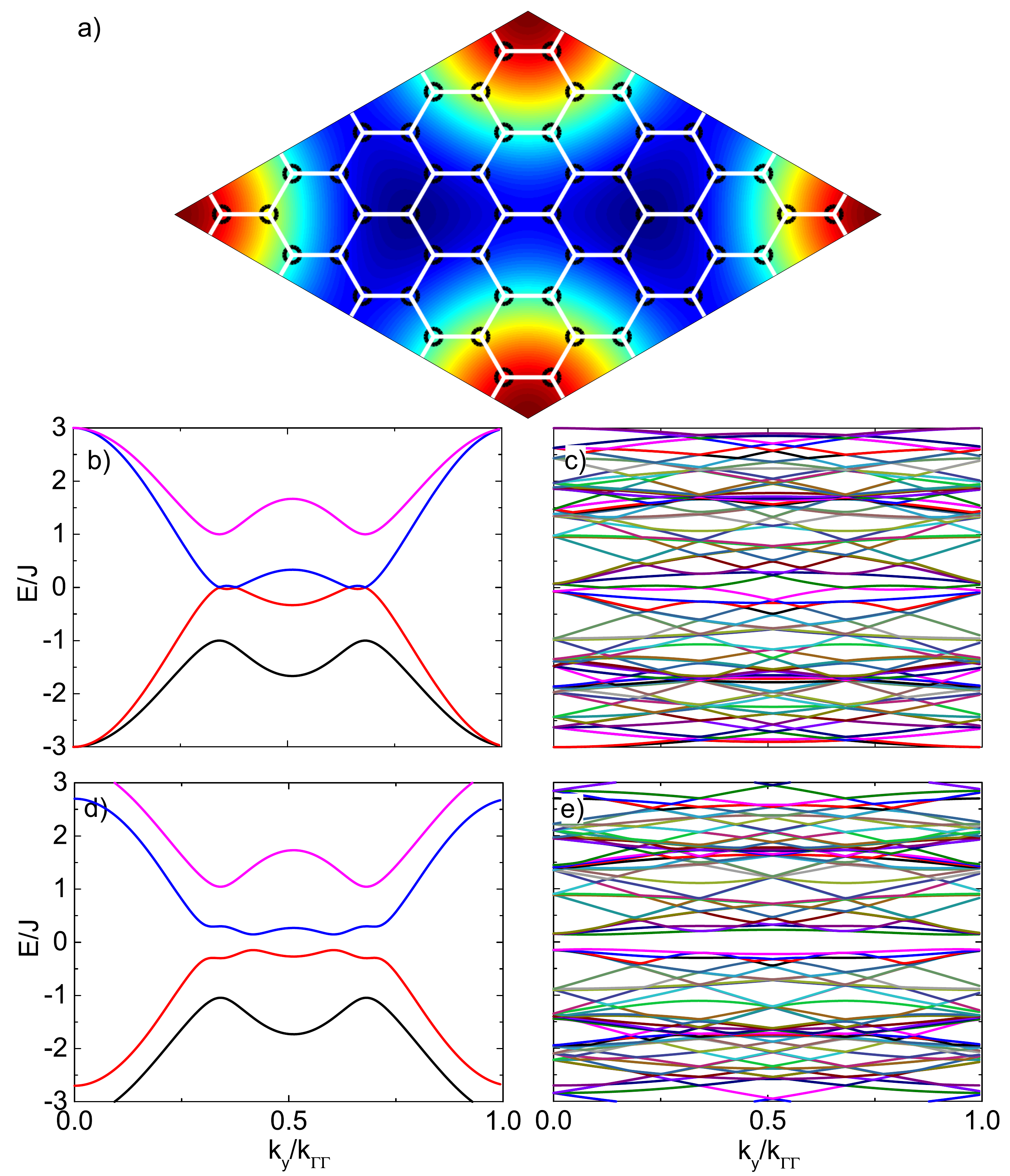}
\caption{Modulated moiré TE-TM SOC and dispersions ($\Gamma K M K'\Gamma$ cut). (a) The moiré unit cell with false color showing the modulation of the TE-TM strength, the white lines represent links between neighboring sites; (b,c) dispersions for $m=1,4$, respectively, with $\delta J=0.45 J$ for trigonal warping visibility ($\Delta=0$); (d,e) dispersions for $m=1,4$ for $\Delta=0.02$.
\label{fig1}}
\end{figure}

Figure~\ref{fig1} illustrates the features of photonic graphene with moiré TE-TM SOC in comparison with constant TE-TM SOC. Fig.~\ref{fig1}(a) shows an example of a moiré unit cell for $m=4$ with the magnitude of the TE-TM SOC shown by color. Circles mark the positions of the atoms, and white lines represent the links between nearest neighbors along which the tunneling occurs.

We first show the example of the dispersion of polariton graphene with TE-TM SOC in Fig.~\ref{fig1}(b) for $\delta J=0.45 J$, making the trigonal warping clearly visible close to the Dirac points $K$ and $K'$ visible in the figure representing the $\Gamma K M K' \Gamma$ cut of the reciprocal space. The effect of the Zeeman splitting $\Delta=0.02 J$ is shown in Fig.~\ref{fig1}(d): a gap opens around $E=0$, with its narrowest points determined by the initial trigonal warping. Next, we show the effect of the modulation of the TE-TM magnitude in Fig.~\ref{fig1}(c). The total number of bands is $m^2$, which means that the average energy width of each band is $6J/m^2$. At the same time, the $\Gamma-\Gamma'$ distance is reduced as $1/m$, meaning that on average, the Fermi velocity given by the slopes at the Dirac points should also decrease as $1/m$. However, it is already visible from the figure that the band density is not uniform, meaning that the width of a given band exhibits a more complicated dependence on $m$ than simply $6J/m^2$. Finally, the effect of the Zeeman splitting on polariton graphene with moiré-modulated TE-TM is shown in Fig.~\ref{fig1}(e). Again, a gap opens around $E=0$, with the bands surrounding this gap exhibiting a visible flattening.

\begin{figure}[tbp]
\centering
\includegraphics[width=0.99\linewidth]{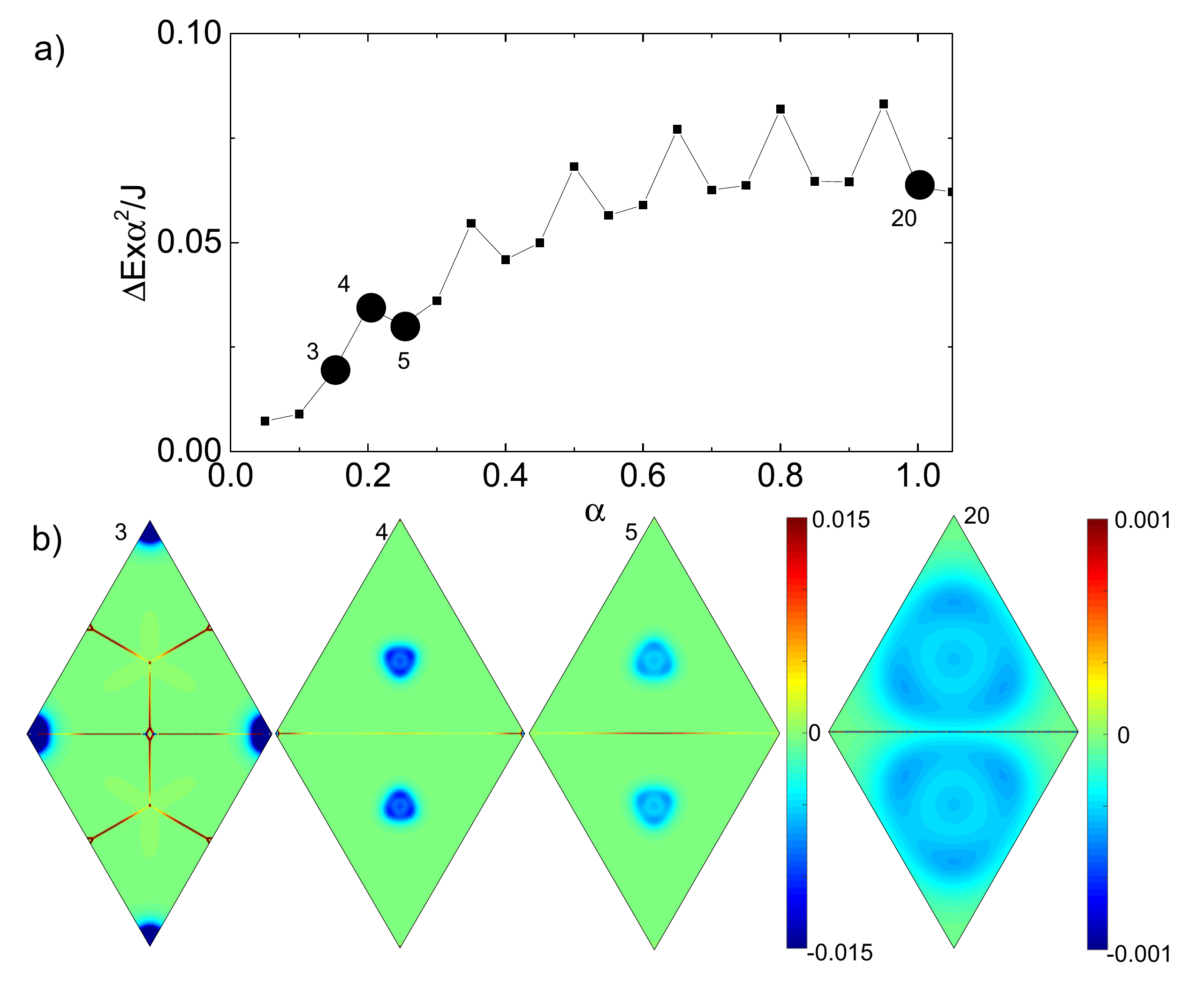}
\caption{a) Normalized band width $\Delta E\times \alpha^2$ (left, black) and Chern numbers (right, red) as a function of moiré parameter $\alpha$; b) Berry curvature for 4 values of $\alpha$, exhibiting periodic symmetry switching.
\label{fig2}}
\end{figure}

In Fig.~\ref{fig2}(a), we plot the width of a particular band multiplied by $m^2$, in order to elucidate the existence of the magic angles, as opposed to the overall $m^{-2}$ decrease of the band width. Following the analogy with twisted bilayer graphene \cite{dosSantos2007,bistritzer2011moire}, the band width is plotted as a function of a dimensionless parameter $\alpha=m\delta J/J$. The local minima are clearly visible and exhibit a periodic pattern. 

To study the topology of our flat bands, we also calculate the Berry curvature and the Chern number $\mathcal{C}$. The distribution of the Berry curvature in the reciprocal space is shown in Fig.~\ref{fig2}(b) (subpanels corresponding to points indicated by numbers in panel (a)). The periodic oscillations of the band width are accompanied by the switching of the symmetry of the Berry curvature distributions. In Fig.~\ref{fig2}(b4,b5,b20)  Berry curvature is localized at the Dirac points in reciprocal space, and the switching of its symmetry between b4 and b5 can be understood from the switching of the local symmetry of the high-symmetry points of the moiré unit cell in real space (see Fig.~S1 in \cite{suppl}). This symmetry also explains the 6-fold symmetry of the Berry curvature distribution in Fig.~\ref{fig2}(b3). The periodicity is illustrated by subpanels b5 and b20, exhibiting the same trigonal distribution, but with a larger scale due to the band flattening.

We note that the topology and the bandwidth exhibit a complicated interdependence. The former is determined by the atom arrangement symmetry and the topological transitions discovered already for non-moiré TE-TM/Zeeman configuration \cite{Nalitov2015b,Bleu2017x}. These transitions are associated with the scale of the trigonal warping (controlled by $\delta J/J$) and the scale of the Zeeman splitting (controlled by $\Delta/J$). In the moiré case, the denominator (the "normal" band width) is reduced by $m^2$. Ultimately, when $m$ is sufficiently large, the topology for a given moiré band becomes trivial (all states become circular-polarized). Before this trivialization occurs, the Berry curvature distribution and the band widths are controlled by the interaction of the bands separated by narrow minigaps (see \cite{suppl}). Because of this, the bands strongly influence each other, and the resulting band Chern numbers can be quite large (for instance $|\mathcal{C}=16|$, see \cite{suppl}). The band width exhibits a regular periodic behavior when the number of closely interacting bands is small, as shown in Fig.~\ref{fig2}(a). This periodicity, determined by the atomic arrangement symmetry, is accompanied by an overall band width decay. In general, we observe that the overall decay becomes faster than $m^2$ for sufficiently large $m$, when the bands start to become trivial, and the critical value of $m$ decreases with $\delta J$ and $\Delta$, because they make the topological transition towards $\mathcal{C}=0$ easier.

\begin{figure}[tbp]
\centering
\includegraphics[width=0.99\linewidth]{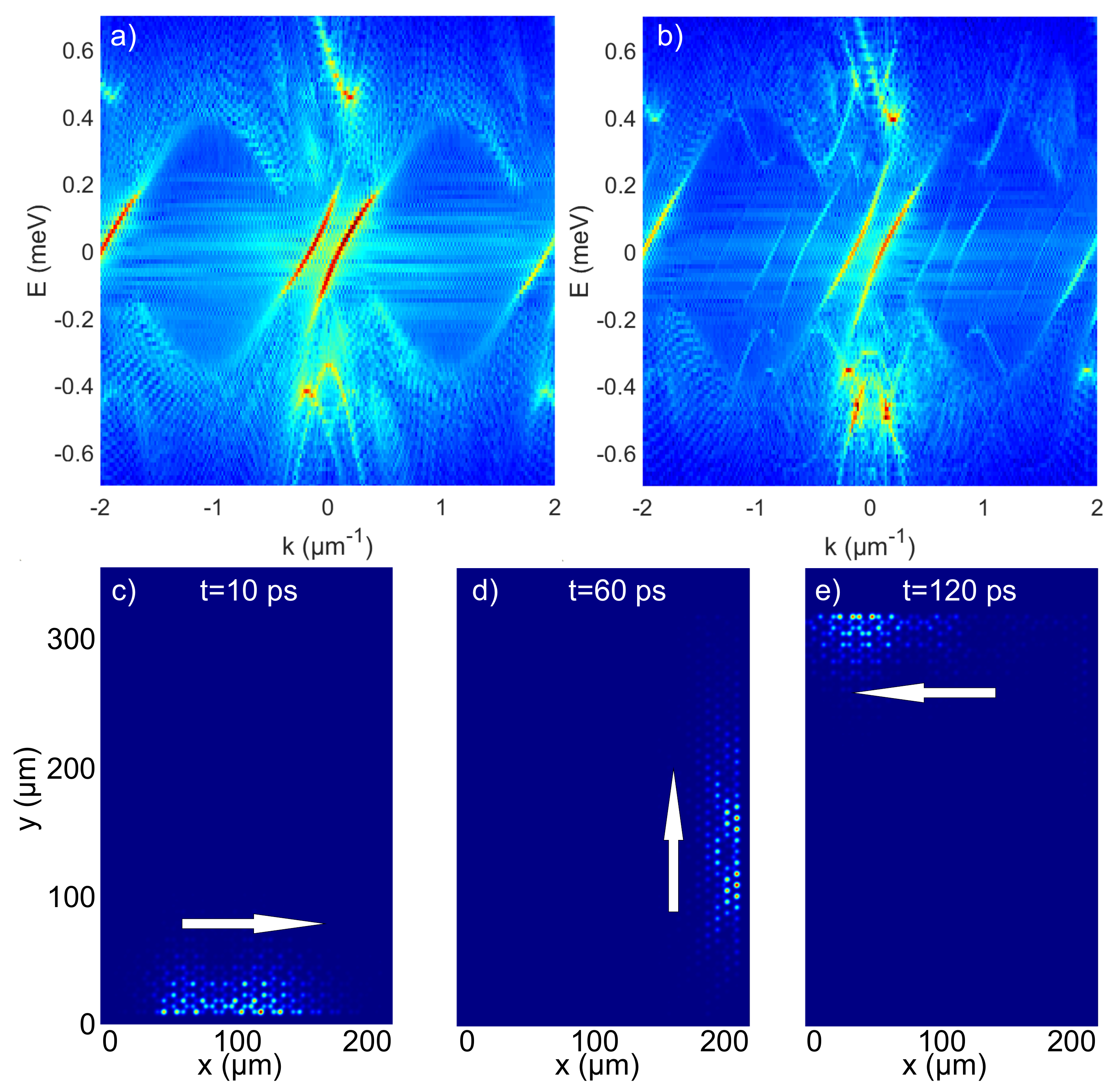}
\caption{(a,b) Dispersion of the honeycomb lattice with TE-TM SOC and Zeeman splitting, showing edge states: (a) $m=1$ (reference), (b) $m=3$, the BZ side length divided by $3$. (c-e) Spatial distribution of the probability density $|\psi(x,y,t)|^2$ from numerical simulations based on Eq.~\eqref{HTBm}: (c) initial ($t=10$~ps), (d) after 1 corner, $t=60$~ps, (e) after 2 corners, $t=120$~ps.
\label{fig3}}
\end{figure}

The topology of the central gap, determined by the sum of the band Chern numbers below the gap $\mathcal{C}=|\sum\mathcal{C}_i|$, is robust with respect to the moiré modulation which does not carry any winding, but it can be modified by transitions associated with trigonal warping ($\mathcal{C}=2\to \mathcal{C}=1$). For a sufficiently small $m$ ($k_0/m>k_{trw}$) and realistic $\delta J$, the gap Chern number remains constant: $\mathcal{C}=2$. The magnitude of the gap also remains the same, determined by the smallest between $\delta J$ and $\Delta$. There is, therefore, a topological band gap of a constant size surrounded by flat topological bands. Topological bulk-boundary correspondence \cite{Hatsugai1993} guarantees the existence of chiral edge states for a finite-sized sample, with their group velocity unchanged. We, therefore, have a system with an increasingly high contrast of group velocity between the flat bulk bands and the chiral propagative edge states. In Fig.~\ref{fig3}(a,b), we show the numerically calculated dispersion of the edge states (with a part of the bulk) for $m=1$ and $m=3$, respectively. The edge states are visible in the middle of the figure (where there is a Dirac point in the absence of a magnetic field). We clearly see the replica of the edge states for $m=3$, appearing at the position of additional Dirac points due to the shrinking of the BZ. Fig.~\ref{fig3}(c-e) present 3 snapshots of the wavepacket propagation along the edge states in a finite-size system, corresponding to 3 different moments. No backscattering occurs at each of the corners of the system, and the propagation velocity is quite high. This confirms the robustness of the topological edge states even in the flat-band moiré systems.

To conclude, we have studied polariton graphene with modulated TE-TM spin-orbit coupling creating a moiré lattice with the unit cell magnification factor $m$. The width of the bands exhibits local minima at periodic values of $m$ corresponding to "magic angles". We demonstrate the switching of the Berry curvature distribution symmetry. Propagative topological edge states, due to their robustness, demonstrate a high contrast with flat bulk bands.

\begin{acknowledgments}
This research was supported by the ANR Labex GaNext (ANR-11-LABX-0014), the ANR program "Investissements d'Avenir" through the IDEX-ISITE initiative 16-IDEX-0001 (CAP 20-25), the ANR projects "NEWAVE" (ANR-21-CE24-0019) and MoirePlusPlus, and the European Union's Horizon 2020 program, through a FET Open research and innovation action under the grant agreement No. 964770 (TopoLight).
\end{acknowledgments}

\bibliography{biblio}

\begin{thebibliography}{55}
\expandafter\ifx\csname natexlab\endcsname\relax\def\natexlab#1{#1}\fi
\expandafter\ifx\csname bibnamefont\endcsname\relax
  \def\bibnamefont#1{#1}\fi
\expandafter\ifx\csname bibfnamefont\endcsname\relax
  \def\bibfnamefont#1{#1}\fi
\expandafter\ifx\csname citenamefont\endcsname\relax
  \def\citenamefont#1{#1}\fi
\expandafter\ifx\csname url\endcsname\relax
  \def\url#1{\texttt{#1}}\fi
\expandafter\ifx\csname urlprefix\endcsname\relax\def\urlprefix{URL }\fi
\providecommand{\bibinfo}[2]{#2}
\providecommand{\eprint}[2][]{\url{#2}}

\bibitem[{\citenamefont{Andrei and MacDonald}(2020)}]{andrei2020graphene}
\bibinfo{author}{\bibfnamefont{E.~Y.} \bibnamefont{Andrei}} \bibnamefont{and}
  \bibinfo{author}{\bibfnamefont{A.~H.} \bibnamefont{MacDonald}},
  \bibinfo{journal}{Nature materials} \textbf{\bibinfo{volume}{19}},
  \bibinfo{pages}{1265} (\bibinfo{year}{2020}).

\bibitem[{\citenamefont{He et~al.}(2021)\citenamefont{He, Zhou, Ye, Cho, Jeong,
  Meng, and Wang}}]{he2021moire}
\bibinfo{author}{\bibfnamefont{F.}~\bibnamefont{He}},
  \bibinfo{author}{\bibfnamefont{Y.}~\bibnamefont{Zhou}},
  \bibinfo{author}{\bibfnamefont{Z.}~\bibnamefont{Ye}},
  \bibinfo{author}{\bibfnamefont{S.-H.} \bibnamefont{Cho}},
  \bibinfo{author}{\bibfnamefont{J.}~\bibnamefont{Jeong}},
  \bibinfo{author}{\bibfnamefont{X.}~\bibnamefont{Meng}}, \bibnamefont{and}
  \bibinfo{author}{\bibfnamefont{Y.}~\bibnamefont{Wang}}, \bibinfo{journal}{ACS
  nano} \textbf{\bibinfo{volume}{15}}, \bibinfo{pages}{5944}
  (\bibinfo{year}{2021}).

\bibitem[{\citenamefont{Huang et~al.}(2022)\citenamefont{Huang, Choi, Shih, and
  Li}}]{huang2022excitons}
\bibinfo{author}{\bibfnamefont{D.}~\bibnamefont{Huang}},
  \bibinfo{author}{\bibfnamefont{J.}~\bibnamefont{Choi}},
  \bibinfo{author}{\bibfnamefont{C.-K.} \bibnamefont{Shih}}, \bibnamefont{and}
  \bibinfo{author}{\bibfnamefont{X.}~\bibnamefont{Li}},
  \bibinfo{journal}{Nature nanotechnology} \textbf{\bibinfo{volume}{17}},
  \bibinfo{pages}{227} (\bibinfo{year}{2022}).

\bibitem[{\citenamefont{Mak and Shan}(2022)}]{mak2022semiconductor}
\bibinfo{author}{\bibfnamefont{K.~F.} \bibnamefont{Mak}} \bibnamefont{and}
  \bibinfo{author}{\bibfnamefont{J.}~\bibnamefont{Shan}},
  \bibinfo{journal}{Nature Nanotechnology} \textbf{\bibinfo{volume}{17}},
  \bibinfo{pages}{686} (\bibinfo{year}{2022}).

\bibitem[{\citenamefont{Du et~al.}(2023)\citenamefont{Du, Molas, Huang, Zhang,
  Wang, and Sun}}]{du2023moire}
\bibinfo{author}{\bibfnamefont{L.}~\bibnamefont{Du}},
  \bibinfo{author}{\bibfnamefont{M.~R.} \bibnamefont{Molas}},
  \bibinfo{author}{\bibfnamefont{Z.}~\bibnamefont{Huang}},
  \bibinfo{author}{\bibfnamefont{G.}~\bibnamefont{Zhang}},
  \bibinfo{author}{\bibfnamefont{F.}~\bibnamefont{Wang}}, \bibnamefont{and}
  \bibinfo{author}{\bibfnamefont{Z.}~\bibnamefont{Sun}},
  \bibinfo{journal}{Science} \textbf{\bibinfo{volume}{379}},
  \bibinfo{pages}{eadg0014} (\bibinfo{year}{2023}).

\bibitem[{\citenamefont{Liu et~al.}(2019)\citenamefont{Liu, Liu, and
  Dai}}]{Liu2019}
\bibinfo{author}{\bibfnamefont{J.}~\bibnamefont{Liu}},
  \bibinfo{author}{\bibfnamefont{J.}~\bibnamefont{Liu}}, \bibnamefont{and}
  \bibinfo{author}{\bibfnamefont{X.}~\bibnamefont{Dai}},
  \bibinfo{journal}{Phys. Rev. B} \textbf{\bibinfo{volume}{99}},
  \bibinfo{pages}{155415} (\bibinfo{year}{2019}),
  \urlprefix\url{https://link.aps.org/doi/10.1103/PhysRevB.99.155415}.

\bibitem[{\citenamefont{Wu et~al.}(2019)\citenamefont{Wu, Lovorn, Tutuc,
  Martin, and MacDonald}}]{wu2019topological}
\bibinfo{author}{\bibfnamefont{F.}~\bibnamefont{Wu}},
  \bibinfo{author}{\bibfnamefont{T.}~\bibnamefont{Lovorn}},
  \bibinfo{author}{\bibfnamefont{E.}~\bibnamefont{Tutuc}},
  \bibinfo{author}{\bibfnamefont{I.}~\bibnamefont{Martin}}, \bibnamefont{and}
  \bibinfo{author}{\bibfnamefont{A.}~\bibnamefont{MacDonald}},
  \bibinfo{journal}{Physical review letters} \textbf{\bibinfo{volume}{122}},
  \bibinfo{pages}{086402} (\bibinfo{year}{2019}).

\bibitem[{\citenamefont{Li et~al.}(2021)\citenamefont{Li, Jiang, Shen, Zhang,
  Li, Tao, Devakul, Watanabe, Taniguchi, Fu et~al.}}]{li2021quantum}
\bibinfo{author}{\bibfnamefont{T.}~\bibnamefont{Li}},
  \bibinfo{author}{\bibfnamefont{S.}~\bibnamefont{Jiang}},
  \bibinfo{author}{\bibfnamefont{B.}~\bibnamefont{Shen}},
  \bibinfo{author}{\bibfnamefont{Y.}~\bibnamefont{Zhang}},
  \bibinfo{author}{\bibfnamefont{L.}~\bibnamefont{Li}},
  \bibinfo{author}{\bibfnamefont{Z.}~\bibnamefont{Tao}},
  \bibinfo{author}{\bibfnamefont{T.}~\bibnamefont{Devakul}},
  \bibinfo{author}{\bibfnamefont{K.}~\bibnamefont{Watanabe}},
  \bibinfo{author}{\bibfnamefont{T.}~\bibnamefont{Taniguchi}},
  \bibinfo{author}{\bibfnamefont{L.}~\bibnamefont{Fu}}, \bibnamefont{et~al.},
  \bibinfo{journal}{Nature} \textbf{\bibinfo{volume}{600}},
  \bibinfo{pages}{641} (\bibinfo{year}{2021}).

\bibitem[{\citenamefont{Cao et~al.}(2018)\citenamefont{Cao, Fatemi, Fang,
  Watanabe, Taniguchi, Kaxiras, and Jarillo-Herrero}}]{Hererro2018}
\bibinfo{author}{\bibfnamefont{Y.}~\bibnamefont{Cao}},
  \bibinfo{author}{\bibfnamefont{V.}~\bibnamefont{Fatemi}},
  \bibinfo{author}{\bibfnamefont{S.}~\bibnamefont{Fang}},
  \bibinfo{author}{\bibfnamefont{K.}~\bibnamefont{Watanabe}},
  \bibinfo{author}{\bibfnamefont{T.}~\bibnamefont{Taniguchi}},
  \bibinfo{author}{\bibfnamefont{E.}~\bibnamefont{Kaxiras}}, \bibnamefont{and}
  \bibinfo{author}{\bibfnamefont{P.}~\bibnamefont{Jarillo-Herrero}},
  \bibinfo{journal}{Nature} \textbf{\bibinfo{volume}{556}}, \bibinfo{pages}{43}
  (\bibinfo{year}{2018}).

\bibitem[{\citenamefont{Liu et~al.}(2012)\citenamefont{Liu, Bergholtz, Fan, and
  L\"auchli}}]{Liu2012}
\bibinfo{author}{\bibfnamefont{Z.}~\bibnamefont{Liu}},
  \bibinfo{author}{\bibfnamefont{E.~J.} \bibnamefont{Bergholtz}},
  \bibinfo{author}{\bibfnamefont{H.}~\bibnamefont{Fan}}, \bibnamefont{and}
  \bibinfo{author}{\bibfnamefont{A.~M.} \bibnamefont{L\"auchli}},
  \bibinfo{journal}{Phys. Rev. Lett.} \textbf{\bibinfo{volume}{109}},
  \bibinfo{pages}{186805} (\bibinfo{year}{2012}),
  \urlprefix\url{https://link.aps.org/doi/10.1103/PhysRevLett.109.186805}.

\bibitem[{\citenamefont{Yang et~al.}(2012)\citenamefont{Yang, Gu, Sun, and
  Das~Sarma}}]{Yang2012}
\bibinfo{author}{\bibfnamefont{S.}~\bibnamefont{Yang}},
  \bibinfo{author}{\bibfnamefont{Z.-C.} \bibnamefont{Gu}},
  \bibinfo{author}{\bibfnamefont{K.}~\bibnamefont{Sun}}, \bibnamefont{and}
  \bibinfo{author}{\bibfnamefont{S.}~\bibnamefont{Das~Sarma}},
  \bibinfo{journal}{Phys. Rev. B} \textbf{\bibinfo{volume}{86}},
  \bibinfo{pages}{241112} (\bibinfo{year}{2012}),
  \urlprefix\url{https://link.aps.org/doi/10.1103/PhysRevB.86.241112}.

\bibitem[{\citenamefont{Bergholtz and Liu}(2013)}]{bergholtz2013topological}
\bibinfo{author}{\bibfnamefont{E.~J.} \bibnamefont{Bergholtz}}
  \bibnamefont{and} \bibinfo{author}{\bibfnamefont{Z.}~\bibnamefont{Liu}},
  \bibinfo{journal}{International Journal of Modern Physics B}
  \textbf{\bibinfo{volume}{27}}, \bibinfo{pages}{1330017}
  (\bibinfo{year}{2013}).

\bibitem[{\citenamefont{Parameswaran et~al.}(2013)\citenamefont{Parameswaran,
  Roy, and Sondhi}}]{Parameswaran2013}
\bibinfo{author}{\bibfnamefont{S.~A.} \bibnamefont{Parameswaran}},
  \bibinfo{author}{\bibfnamefont{R.}~\bibnamefont{Roy}}, \bibnamefont{and}
  \bibinfo{author}{\bibfnamefont{S.~L.} \bibnamefont{Sondhi}},
  \bibinfo{journal}{Comptes Rendus Physique} \textbf{\bibinfo{volume}{14}},
  \bibinfo{pages}{816} (\bibinfo{year}{2013}), ISSN \bibinfo{issn}{1631-0705},
  \bibinfo{note}{topological insulators / Isolants topologiques},
  \urlprefix\url{https://www.sciencedirect.com/science/article/pii/S163107051300073X}.

\bibitem[{\citenamefont{Wu et~al.}(2021)\citenamefont{Wu, Zhang, Watanabe,
  Taniguchi, and Andrei}}]{wu2021chern}
\bibinfo{author}{\bibfnamefont{S.}~\bibnamefont{Wu}},
  \bibinfo{author}{\bibfnamefont{Z.}~\bibnamefont{Zhang}},
  \bibinfo{author}{\bibfnamefont{K.}~\bibnamefont{Watanabe}},
  \bibinfo{author}{\bibfnamefont{T.}~\bibnamefont{Taniguchi}},
  \bibnamefont{and} \bibinfo{author}{\bibfnamefont{E.~Y.}
  \bibnamefont{Andrei}}, \bibinfo{journal}{Nature materials}
  \textbf{\bibinfo{volume}{20}}, \bibinfo{pages}{488} (\bibinfo{year}{2021}).

\bibitem[{\citenamefont{Bissbort et~al.}(2013)\citenamefont{Bissbort, Cocks,
  Negretti, Idziaszek, Calarco, Schmidt-Kaler, Hofstetter, and
  Gerritsma}}]{Bissbort2013}
\bibinfo{author}{\bibfnamefont{U.}~\bibnamefont{Bissbort}},
  \bibinfo{author}{\bibfnamefont{D.}~\bibnamefont{Cocks}},
  \bibinfo{author}{\bibfnamefont{A.}~\bibnamefont{Negretti}},
  \bibinfo{author}{\bibfnamefont{Z.}~\bibnamefont{Idziaszek}},
  \bibinfo{author}{\bibfnamefont{T.}~\bibnamefont{Calarco}},
  \bibinfo{author}{\bibfnamefont{F.}~\bibnamefont{Schmidt-Kaler}},
  \bibinfo{author}{\bibfnamefont{W.}~\bibnamefont{Hofstetter}},
  \bibnamefont{and}
  \bibinfo{author}{\bibfnamefont{R.}~\bibnamefont{Gerritsma}},
  \bibinfo{journal}{Phys. Rev. Lett.} \textbf{\bibinfo{volume}{111}},
  \bibinfo{pages}{080501} (\bibinfo{year}{2013}),
  \urlprefix\url{https://link.aps.org/doi/10.1103/PhysRevLett.111.080501}.

\bibitem[{\citenamefont{Gross and Bloch}(2017)}]{gross2017quantum}
\bibinfo{author}{\bibfnamefont{C.}~\bibnamefont{Gross}} \bibnamefont{and}
  \bibinfo{author}{\bibfnamefont{I.}~\bibnamefont{Bloch}},
  \bibinfo{journal}{Science} \textbf{\bibinfo{volume}{357}},
  \bibinfo{pages}{995} (\bibinfo{year}{2017}).

\bibitem[{\citenamefont{He et~al.}(2016)\citenamefont{He, Ni, Ge, Sun, Chen,
  Lu, Liu, and Chen}}]{he2016acoustic}
\bibinfo{author}{\bibfnamefont{C.}~\bibnamefont{He}},
  \bibinfo{author}{\bibfnamefont{X.}~\bibnamefont{Ni}},
  \bibinfo{author}{\bibfnamefont{H.}~\bibnamefont{Ge}},
  \bibinfo{author}{\bibfnamefont{X.-C.} \bibnamefont{Sun}},
  \bibinfo{author}{\bibfnamefont{Y.-B.} \bibnamefont{Chen}},
  \bibinfo{author}{\bibfnamefont{M.-H.} \bibnamefont{Lu}},
  \bibinfo{author}{\bibfnamefont{X.-P.} \bibnamefont{Liu}}, \bibnamefont{and}
  \bibinfo{author}{\bibfnamefont{Y.-F.} \bibnamefont{Chen}},
  \bibinfo{journal}{Nature physics} \textbf{\bibinfo{volume}{12}},
  \bibinfo{pages}{1124} (\bibinfo{year}{2016}).

\bibitem[{\citenamefont{Ozawa et~al.}(2019)\citenamefont{Ozawa, Price, Amo,
  Goldman, Hafezi, Lu, Rechtsman, Schuster, Simon, Zilberberg
  et~al.}}]{Ozawa2019}
\bibinfo{author}{\bibfnamefont{T.}~\bibnamefont{Ozawa}},
  \bibinfo{author}{\bibfnamefont{H.~M.} \bibnamefont{Price}},
  \bibinfo{author}{\bibfnamefont{A.}~\bibnamefont{Amo}},
  \bibinfo{author}{\bibfnamefont{N.}~\bibnamefont{Goldman}},
  \bibinfo{author}{\bibfnamefont{M.}~\bibnamefont{Hafezi}},
  \bibinfo{author}{\bibfnamefont{L.}~\bibnamefont{Lu}},
  \bibinfo{author}{\bibfnamefont{M.~C.} \bibnamefont{Rechtsman}},
  \bibinfo{author}{\bibfnamefont{D.}~\bibnamefont{Schuster}},
  \bibinfo{author}{\bibfnamefont{J.}~\bibnamefont{Simon}},
  \bibinfo{author}{\bibfnamefont{O.}~\bibnamefont{Zilberberg}},
  \bibnamefont{et~al.}, \bibinfo{journal}{Rev. Mod. Phys.}
  \textbf{\bibinfo{volume}{91}}, \bibinfo{pages}{015006}
  (\bibinfo{year}{2019}),
  \urlprefix\url{https://link.aps.org/doi/10.1103/RevModPhys.91.015006}.

\bibitem[{\citenamefont{Xue et~al.}(2022)\citenamefont{Xue, Yang, and
  Zhang}}]{xue2022topological}
\bibinfo{author}{\bibfnamefont{H.}~\bibnamefont{Xue}},
  \bibinfo{author}{\bibfnamefont{Y.}~\bibnamefont{Yang}}, \bibnamefont{and}
  \bibinfo{author}{\bibfnamefont{B.}~\bibnamefont{Zhang}},
  \bibinfo{journal}{Nature Reviews Materials} \textbf{\bibinfo{volume}{7}},
  \bibinfo{pages}{974} (\bibinfo{year}{2022}).

\bibitem[{\citenamefont{Wang et~al.}(2020{\natexlab{a}})\citenamefont{Wang,
  Price, Zhang, and Chong}}]{wang2020circuit}
\bibinfo{author}{\bibfnamefont{Y.}~\bibnamefont{Wang}},
  \bibinfo{author}{\bibfnamefont{H.~M.} \bibnamefont{Price}},
  \bibinfo{author}{\bibfnamefont{B.}~\bibnamefont{Zhang}}, \bibnamefont{and}
  \bibinfo{author}{\bibfnamefont{Y.}~\bibnamefont{Chong}},
  \bibinfo{journal}{Nature communications} \textbf{\bibinfo{volume}{11}},
  \bibinfo{pages}{2356} (\bibinfo{year}{2020}{\natexlab{a}}).

\bibitem[{\citenamefont{Mao et~al.}(2021)\citenamefont{Mao, Shao, Luan, Wang,
  and Ma}}]{mao2021magic}
\bibinfo{author}{\bibfnamefont{X.-R.} \bibnamefont{Mao}},
  \bibinfo{author}{\bibfnamefont{Z.-K.} \bibnamefont{Shao}},
  \bibinfo{author}{\bibfnamefont{H.-Y.} \bibnamefont{Luan}},
  \bibinfo{author}{\bibfnamefont{S.-L.} \bibnamefont{Wang}}, \bibnamefont{and}
  \bibinfo{author}{\bibfnamefont{R.-M.} \bibnamefont{Ma}},
  \bibinfo{journal}{Nature nanotechnology} \textbf{\bibinfo{volume}{16}},
  \bibinfo{pages}{1099} (\bibinfo{year}{2021}).

\bibitem[{\citenamefont{Wang et~al.}(2020{\natexlab{b}})\citenamefont{Wang,
  Zheng, Chen, Huang, Kartashov, Torner, Konotop, and
  Ye}}]{wang2020localization}
\bibinfo{author}{\bibfnamefont{P.}~\bibnamefont{Wang}},
  \bibinfo{author}{\bibfnamefont{Y.}~\bibnamefont{Zheng}},
  \bibinfo{author}{\bibfnamefont{X.}~\bibnamefont{Chen}},
  \bibinfo{author}{\bibfnamefont{C.}~\bibnamefont{Huang}},
  \bibinfo{author}{\bibfnamefont{Y.~V.} \bibnamefont{Kartashov}},
  \bibinfo{author}{\bibfnamefont{L.}~\bibnamefont{Torner}},
  \bibinfo{author}{\bibfnamefont{V.~V.} \bibnamefont{Konotop}},
  \bibnamefont{and} \bibinfo{author}{\bibfnamefont{F.}~\bibnamefont{Ye}},
  \bibinfo{journal}{Nature} \textbf{\bibinfo{volume}{577}}, \bibinfo{pages}{42}
  (\bibinfo{year}{2020}{\natexlab{b}}).

\bibitem[{\citenamefont{Wang et~al.}(2022)\citenamefont{Wang, Fu, Peng,
  Kartashov, Torner, Konotop, and Ye}}]{wang2022two}
\bibinfo{author}{\bibfnamefont{P.}~\bibnamefont{Wang}},
  \bibinfo{author}{\bibfnamefont{Q.}~\bibnamefont{Fu}},
  \bibinfo{author}{\bibfnamefont{R.}~\bibnamefont{Peng}},
  \bibinfo{author}{\bibfnamefont{Y.~V.} \bibnamefont{Kartashov}},
  \bibinfo{author}{\bibfnamefont{L.}~\bibnamefont{Torner}},
  \bibinfo{author}{\bibfnamefont{V.~V.} \bibnamefont{Konotop}},
  \bibnamefont{and} \bibinfo{author}{\bibfnamefont{F.}~\bibnamefont{Ye}},
  \bibinfo{journal}{Nature communications} \textbf{\bibinfo{volume}{13}},
  \bibinfo{pages}{6738} (\bibinfo{year}{2022}).

\bibitem[{\citenamefont{Bajoni et~al.}(2008)\citenamefont{Bajoni, Senellart,
  Wertz, Sagnes, Miard, Lema\^{\i}tre, and Bloch}}]{Bajoni2008}
\bibinfo{author}{\bibfnamefont{D.}~\bibnamefont{Bajoni}},
  \bibinfo{author}{\bibfnamefont{P.}~\bibnamefont{Senellart}},
  \bibinfo{author}{\bibfnamefont{E.}~\bibnamefont{Wertz}},
  \bibinfo{author}{\bibfnamefont{I.}~\bibnamefont{Sagnes}},
  \bibinfo{author}{\bibfnamefont{A.}~\bibnamefont{Miard}},
  \bibinfo{author}{\bibfnamefont{A.}~\bibnamefont{Lema\^{\i}tre}},
  \bibnamefont{and} \bibinfo{author}{\bibfnamefont{J.}~\bibnamefont{Bloch}},
  \bibinfo{journal}{Phys. Rev. Lett.} \textbf{\bibinfo{volume}{100}},
  \bibinfo{pages}{047401} (\bibinfo{year}{2008}),
  \urlprefix\url{https://link.aps.org/doi/10.1103/PhysRevLett.100.047401}.

\bibitem[{\citenamefont{Jacqmin et~al.}(2014)\citenamefont{Jacqmin, Carusotto,
  Sagnes, Abbarchi, Solnyshkov, Malpuech, Galopin, Lema\^{\i}tre, Bloch, and
  Amo}}]{Jacqmin2014}
\bibinfo{author}{\bibfnamefont{T.}~\bibnamefont{Jacqmin}},
  \bibinfo{author}{\bibfnamefont{I.}~\bibnamefont{Carusotto}},
  \bibinfo{author}{\bibfnamefont{I.}~\bibnamefont{Sagnes}},
  \bibinfo{author}{\bibfnamefont{M.}~\bibnamefont{Abbarchi}},
  \bibinfo{author}{\bibfnamefont{D.~D.} \bibnamefont{Solnyshkov}},
  \bibinfo{author}{\bibfnamefont{G.}~\bibnamefont{Malpuech}},
  \bibinfo{author}{\bibfnamefont{E.}~\bibnamefont{Galopin}},
  \bibinfo{author}{\bibfnamefont{A.}~\bibnamefont{Lema\^{\i}tre}},
  \bibinfo{author}{\bibfnamefont{J.}~\bibnamefont{Bloch}}, \bibnamefont{and}
  \bibinfo{author}{\bibfnamefont{A.}~\bibnamefont{Amo}},
  \bibinfo{journal}{Phys. Rev. Lett.} \textbf{\bibinfo{volume}{112}},
  \bibinfo{pages}{116402} (\bibinfo{year}{2014}),
  \urlprefix\url{https://link.aps.org/doi/10.1103/PhysRevLett.112.116402}.

\bibitem[{\citenamefont{Whittaker et~al.}(2018)\citenamefont{Whittaker,
  Cancellieri, Walker, Gulevich, Schomerus, Vaitiekus, Royall, Whittaker,
  Clarke, Iorsh et~al.}}]{Whittaker2018}
\bibinfo{author}{\bibfnamefont{C.~E.} \bibnamefont{Whittaker}},
  \bibinfo{author}{\bibfnamefont{E.}~\bibnamefont{Cancellieri}},
  \bibinfo{author}{\bibfnamefont{P.~M.} \bibnamefont{Walker}},
  \bibinfo{author}{\bibfnamefont{D.~R.} \bibnamefont{Gulevich}},
  \bibinfo{author}{\bibfnamefont{H.}~\bibnamefont{Schomerus}},
  \bibinfo{author}{\bibfnamefont{D.}~\bibnamefont{Vaitiekus}},
  \bibinfo{author}{\bibfnamefont{B.}~\bibnamefont{Royall}},
  \bibinfo{author}{\bibfnamefont{D.~M.} \bibnamefont{Whittaker}},
  \bibinfo{author}{\bibfnamefont{E.}~\bibnamefont{Clarke}},
  \bibinfo{author}{\bibfnamefont{I.~V.} \bibnamefont{Iorsh}},
  \bibnamefont{et~al.}, \bibinfo{journal}{Phys. Rev. Lett.}
  \textbf{\bibinfo{volume}{120}}, \bibinfo{pages}{097401}
  (\bibinfo{year}{2018}),
  \urlprefix\url{https://link.aps.org/doi/10.1103/PhysRevLett.120.097401}.

\bibitem[{\citenamefont{Klembt et~al.}(2018)\citenamefont{Klembt, Harder,
  Egorov, Winkler, Ge, Bandres, Emmerling, Worschech, Liew, Segev
  et~al.}}]{klembt2018exciton}
\bibinfo{author}{\bibfnamefont{S.}~\bibnamefont{Klembt}},
  \bibinfo{author}{\bibfnamefont{T.}~\bibnamefont{Harder}},
  \bibinfo{author}{\bibfnamefont{O.}~\bibnamefont{Egorov}},
  \bibinfo{author}{\bibfnamefont{K.}~\bibnamefont{Winkler}},
  \bibinfo{author}{\bibfnamefont{R.}~\bibnamefont{Ge}},
  \bibinfo{author}{\bibfnamefont{M.}~\bibnamefont{Bandres}},
  \bibinfo{author}{\bibfnamefont{M.}~\bibnamefont{Emmerling}},
  \bibinfo{author}{\bibfnamefont{L.}~\bibnamefont{Worschech}},
  \bibinfo{author}{\bibfnamefont{T.}~\bibnamefont{Liew}},
  \bibinfo{author}{\bibfnamefont{M.}~\bibnamefont{Segev}},
  \bibnamefont{et~al.}, \bibinfo{journal}{Nature}
  \textbf{\bibinfo{volume}{562}}, \bibinfo{pages}{552} (\bibinfo{year}{2018}).

\bibitem[{\citenamefont{Gianfrate et~al.}(2020)\citenamefont{Gianfrate, Bleu,
  Dominici, Ardizzone, De~Giorgi, Ballarini, Lerario, West, Pfeiffer,
  Solnyshkov et~al.}}]{gianfrate2020measurement}
\bibinfo{author}{\bibfnamefont{A.}~\bibnamefont{Gianfrate}},
  \bibinfo{author}{\bibfnamefont{O.}~\bibnamefont{Bleu}},
  \bibinfo{author}{\bibfnamefont{L.}~\bibnamefont{Dominici}},
  \bibinfo{author}{\bibfnamefont{V.}~\bibnamefont{Ardizzone}},
  \bibinfo{author}{\bibfnamefont{M.}~\bibnamefont{De~Giorgi}},
  \bibinfo{author}{\bibfnamefont{D.}~\bibnamefont{Ballarini}},
  \bibinfo{author}{\bibfnamefont{G.}~\bibnamefont{Lerario}},
  \bibinfo{author}{\bibfnamefont{K.}~\bibnamefont{West}},
  \bibinfo{author}{\bibfnamefont{L.}~\bibnamefont{Pfeiffer}},
  \bibinfo{author}{\bibfnamefont{D.}~\bibnamefont{Solnyshkov}},
  \bibnamefont{et~al.}, \bibinfo{journal}{Nature}
  \textbf{\bibinfo{volume}{578}}, \bibinfo{pages}{381} (\bibinfo{year}{2020}).

\bibitem[{\citenamefont{Kavokin et~al.}(2011)\citenamefont{Kavokin, Baumberg,
  Malpuech, and Laussy}}]{Microcavities}
\bibinfo{author}{\bibfnamefont{A.}~\bibnamefont{Kavokin}},
  \bibinfo{author}{\bibfnamefont{J.~J.} \bibnamefont{Baumberg}},
  \bibinfo{author}{\bibfnamefont{G.}~\bibnamefont{Malpuech}}, \bibnamefont{and}
  \bibinfo{author}{\bibfnamefont{F.~P.} \bibnamefont{Laussy}},
  \emph{\bibinfo{title}{Microcavities}} (\bibinfo{publisher}{Oxford University
  Press}, \bibinfo{year}{2011}).

\bibitem[{\citenamefont{Solnyshkov et~al.}(2016)\citenamefont{Solnyshkov,
  Nalitov, and Malpuech}}]{Solnyshkov2016}
\bibinfo{author}{\bibfnamefont{D.~D.} \bibnamefont{Solnyshkov}},
  \bibinfo{author}{\bibfnamefont{A.~V.} \bibnamefont{Nalitov}},
  \bibnamefont{and} \bibinfo{author}{\bibfnamefont{G.}~\bibnamefont{Malpuech}},
  \bibinfo{journal}{Physical Review Letters} \textbf{\bibinfo{volume}{116}},
  \bibinfo{pages}{046402} (\bibinfo{year}{2016}), ISSN
  \bibinfo{issn}{0031-9007},
  \urlprefix\url{http://link.aps.org/doi/10.1103/PhysRevLett.116.046402}.

\bibitem[{\citenamefont{St-Jean et~al.}(2017)\citenamefont{St-Jean, Goblot,
  Galopin, Lemaitre, Ozawa, Le~Gratiet, Sagnes, Bloch, and Amo}}]{StJean2017}
\bibinfo{author}{\bibfnamefont{P.}~\bibnamefont{St-Jean}},
  \bibinfo{author}{\bibfnamefont{V.}~\bibnamefont{Goblot}},
  \bibinfo{author}{\bibfnamefont{E.}~\bibnamefont{Galopin}},
  \bibinfo{author}{\bibfnamefont{A.}~\bibnamefont{Lemaitre}},
  \bibinfo{author}{\bibfnamefont{T.}~\bibnamefont{Ozawa}},
  \bibinfo{author}{\bibfnamefont{L.}~\bibnamefont{Le~Gratiet}},
  \bibinfo{author}{\bibfnamefont{I.}~\bibnamefont{Sagnes}},
  \bibinfo{author}{\bibfnamefont{J.}~\bibnamefont{Bloch}}, \bibnamefont{and}
  \bibinfo{author}{\bibfnamefont{A.}~\bibnamefont{Amo}},
  \bibinfo{journal}{Nature Photonics} \textbf{\bibinfo{volume}{11}},
  \bibinfo{pages}{651} (\bibinfo{year}{2017}).

\bibitem[{\citenamefont{Su et~al.}(1979)\citenamefont{Su, Schrieffer, and
  Heeger}}]{Su1979}
\bibinfo{author}{\bibfnamefont{W.~P.} \bibnamefont{Su}},
  \bibinfo{author}{\bibfnamefont{J.~R.} \bibnamefont{Schrieffer}},
  \bibnamefont{and} \bibinfo{author}{\bibfnamefont{A.~J.}
  \bibnamefont{Heeger}}, \bibinfo{journal}{Phys. Rev. Lett.}
  \textbf{\bibinfo{volume}{42}}, \bibinfo{pages}{1698} (\bibinfo{year}{1979}),
  \urlprefix\url{https://link.aps.org/doi/10.1103/PhysRevLett.42.1698}.

\bibitem[{\citenamefont{Carusotto and Ciuti}(2013)}]{carusotto2013quantum}
\bibinfo{author}{\bibfnamefont{I.}~\bibnamefont{Carusotto}} \bibnamefont{and}
  \bibinfo{author}{\bibfnamefont{C.}~\bibnamefont{Ciuti}},
  \bibinfo{journal}{Reviews of Modern Physics} \textbf{\bibinfo{volume}{85}},
  \bibinfo{pages}{299} (\bibinfo{year}{2013}).

\bibitem[{\citenamefont{Solnyshkov et~al.}(2021)\citenamefont{Solnyshkov,
  Malpuech, St-Jean, Ravets, Bloch, and Amo}}]{solnyshkov2021microcavity}
\bibinfo{author}{\bibfnamefont{D.~D.} \bibnamefont{Solnyshkov}},
  \bibinfo{author}{\bibfnamefont{G.}~\bibnamefont{Malpuech}},
  \bibinfo{author}{\bibfnamefont{P.}~\bibnamefont{St-Jean}},
  \bibinfo{author}{\bibfnamefont{S.}~\bibnamefont{Ravets}},
  \bibinfo{author}{\bibfnamefont{J.}~\bibnamefont{Bloch}}, \bibnamefont{and}
  \bibinfo{author}{\bibfnamefont{A.}~\bibnamefont{Amo}},
  \bibinfo{journal}{Optical Materials Express} \textbf{\bibinfo{volume}{11}},
  \bibinfo{pages}{1119} (\bibinfo{year}{2021}).

\bibitem[{\citenamefont{Mu{\~n}oz-Matutano
  et~al.}(2019)\citenamefont{Mu{\~n}oz-Matutano, Wood, Johnsson, Vidal,
  Baragiola, Reinhard, Lema{\^\i}tre, Bloch, Amo, Nogues
  et~al.}}]{munoz2019emergence}
\bibinfo{author}{\bibfnamefont{G.}~\bibnamefont{Mu{\~n}oz-Matutano}},
  \bibinfo{author}{\bibfnamefont{A.}~\bibnamefont{Wood}},
  \bibinfo{author}{\bibfnamefont{M.}~\bibnamefont{Johnsson}},
  \bibinfo{author}{\bibfnamefont{X.}~\bibnamefont{Vidal}},
  \bibinfo{author}{\bibfnamefont{B.~Q.} \bibnamefont{Baragiola}},
  \bibinfo{author}{\bibfnamefont{A.}~\bibnamefont{Reinhard}},
  \bibinfo{author}{\bibfnamefont{A.}~\bibnamefont{Lema{\^\i}tre}},
  \bibinfo{author}{\bibfnamefont{J.}~\bibnamefont{Bloch}},
  \bibinfo{author}{\bibfnamefont{A.}~\bibnamefont{Amo}},
  \bibinfo{author}{\bibfnamefont{G.}~\bibnamefont{Nogues}},
  \bibnamefont{et~al.}, \bibinfo{journal}{Nature materials}
  \textbf{\bibinfo{volume}{18}}, \bibinfo{pages}{213} (\bibinfo{year}{2019}).

\bibitem[{\citenamefont{Zhang et~al.}(2021)\citenamefont{Zhang, Wu, Hou, Zhang,
  Chou, Watanabe, Taniguchi, Forrest, and Deng}}]{zhang2021van}
\bibinfo{author}{\bibfnamefont{L.}~\bibnamefont{Zhang}},
  \bibinfo{author}{\bibfnamefont{F.}~\bibnamefont{Wu}},
  \bibinfo{author}{\bibfnamefont{S.}~\bibnamefont{Hou}},
  \bibinfo{author}{\bibfnamefont{Z.}~\bibnamefont{Zhang}},
  \bibinfo{author}{\bibfnamefont{Y.-H.} \bibnamefont{Chou}},
  \bibinfo{author}{\bibfnamefont{K.}~\bibnamefont{Watanabe}},
  \bibinfo{author}{\bibfnamefont{T.}~\bibnamefont{Taniguchi}},
  \bibinfo{author}{\bibfnamefont{S.~R.} \bibnamefont{Forrest}},
  \bibnamefont{and} \bibinfo{author}{\bibfnamefont{H.}~\bibnamefont{Deng}},
  \bibinfo{journal}{Nature} \textbf{\bibinfo{volume}{591}}, \bibinfo{pages}{61}
  (\bibinfo{year}{2021}).

\bibitem[{\citenamefont{Zasedatelev et~al.}(2021)\citenamefont{Zasedatelev,
  Baranikov, Sannikov, Urbonas, Scafirimuto, Shishkov, Andrianov, Lozovik,
  Scherf, St{\"o}ferle et~al.}}]{zasedatelev2021single}
\bibinfo{author}{\bibfnamefont{A.~V.} \bibnamefont{Zasedatelev}},
  \bibinfo{author}{\bibfnamefont{A.~V.} \bibnamefont{Baranikov}},
  \bibinfo{author}{\bibfnamefont{D.}~\bibnamefont{Sannikov}},
  \bibinfo{author}{\bibfnamefont{D.}~\bibnamefont{Urbonas}},
  \bibinfo{author}{\bibfnamefont{F.}~\bibnamefont{Scafirimuto}},
  \bibinfo{author}{\bibfnamefont{V.~Y.} \bibnamefont{Shishkov}},
  \bibinfo{author}{\bibfnamefont{E.~S.} \bibnamefont{Andrianov}},
  \bibinfo{author}{\bibfnamefont{Y.~E.} \bibnamefont{Lozovik}},
  \bibinfo{author}{\bibfnamefont{U.}~\bibnamefont{Scherf}},
  \bibinfo{author}{\bibfnamefont{T.}~\bibnamefont{St{\"o}ferle}},
  \bibnamefont{et~al.}, \bibinfo{journal}{Nature}
  \textbf{\bibinfo{volume}{597}}, \bibinfo{pages}{493} (\bibinfo{year}{2021}).

\bibitem[{\citenamefont{Kavokin et~al.}(2005)\citenamefont{Kavokin, Malpuech,
  and Glazov}}]{Kavokin2005}
\bibinfo{author}{\bibfnamefont{A.}~\bibnamefont{Kavokin}},
  \bibinfo{author}{\bibfnamefont{G.}~\bibnamefont{Malpuech}}, \bibnamefont{and}
  \bibinfo{author}{\bibfnamefont{M.}~\bibnamefont{Glazov}},
  \bibinfo{journal}{Phys. Rev. Lett.} \textbf{\bibinfo{volume}{95}},
  \bibinfo{pages}{136601} (\bibinfo{year}{2005}),
  \urlprefix\url{http://link.aps.org/doi/10.1103/PhysRevLett.95.136601}.

\bibitem[{\citenamefont{Ter\ifmmode~\mbox{\c{c}}\else \c{c}\fi{}as
  et~al.}(2014)\citenamefont{Ter\ifmmode~\mbox{\c{c}}\else \c{c}\fi{}as,
  Flayac, Solnyshkov, and Malpuech}}]{Tercas2014}
\bibinfo{author}{\bibfnamefont{H.}~\bibnamefont{Ter\ifmmode~\mbox{\c{c}}\else
  \c{c}\fi{}as}}, \bibinfo{author}{\bibfnamefont{H.}~\bibnamefont{Flayac}},
  \bibinfo{author}{\bibfnamefont{D.~D.} \bibnamefont{Solnyshkov}},
  \bibnamefont{and} \bibinfo{author}{\bibfnamefont{G.}~\bibnamefont{Malpuech}},
  \bibinfo{journal}{Phys. Rev. Lett.} \textbf{\bibinfo{volume}{112}},
  \bibinfo{pages}{066402} (\bibinfo{year}{2014}),
  \urlprefix\url{https://link.aps.org/doi/10.1103/PhysRevLett.112.066402}.

\bibitem[{\citenamefont{Rechci{\'n}ska
  et~al.}(2019)\citenamefont{Rechci{\'n}ska, Kr{\'o}l, Mazur, Morawiak, Mirek,
  {\L}empicka, Bardyszewski, Matuszewski, Kula, Piecek
  et~al.}}]{rechcinska2019engineering}
\bibinfo{author}{\bibfnamefont{K.}~\bibnamefont{Rechci{\'n}ska}},
  \bibinfo{author}{\bibfnamefont{M.}~\bibnamefont{Kr{\'o}l}},
  \bibinfo{author}{\bibfnamefont{R.}~\bibnamefont{Mazur}},
  \bibinfo{author}{\bibfnamefont{P.}~\bibnamefont{Morawiak}},
  \bibinfo{author}{\bibfnamefont{R.}~\bibnamefont{Mirek}},
  \bibinfo{author}{\bibfnamefont{K.}~\bibnamefont{{\L}empicka}},
  \bibinfo{author}{\bibfnamefont{W.}~\bibnamefont{Bardyszewski}},
  \bibinfo{author}{\bibfnamefont{M.}~\bibnamefont{Matuszewski}},
  \bibinfo{author}{\bibfnamefont{P.}~\bibnamefont{Kula}},
  \bibinfo{author}{\bibfnamefont{W.}~\bibnamefont{Piecek}},
  \bibnamefont{et~al.}, \bibinfo{journal}{Science}
  \textbf{\bibinfo{volume}{366}}, \bibinfo{pages}{727} (\bibinfo{year}{2019}).

\bibitem[{\citenamefont{Jiahuan et~al.}(2021)\citenamefont{Jiahuan, Qing, Feng,
  Yiming, Olivier, Guillaume, Jiannian, Hongbing, and Dmitry}}]{Ren2021}
\bibinfo{author}{\bibfnamefont{R.}~\bibnamefont{Jiahuan}},
  \bibinfo{author}{\bibfnamefont{L.}~\bibnamefont{Qing}},
  \bibinfo{author}{\bibfnamefont{L.}~\bibnamefont{Feng}},
  \bibinfo{author}{\bibfnamefont{L.}~\bibnamefont{Yiming}},
  \bibinfo{author}{\bibfnamefont{B.}~\bibnamefont{Olivier}},
  \bibinfo{author}{\bibfnamefont{M.}~\bibnamefont{Guillaume}},
  \bibinfo{author}{\bibfnamefont{Y.}~\bibnamefont{Jiannian}},
  \bibinfo{author}{\bibfnamefont{F.}~\bibnamefont{Hongbing}}, \bibnamefont{and}
  \bibinfo{author}{\bibfnamefont{S.}~\bibnamefont{Dmitry}},
  \bibinfo{journal}{Nature Communications} \textbf{\bibinfo{volume}{12}},
  \bibinfo{pages}{1} (\bibinfo{year}{2021}).

\bibitem[{\citenamefont{Nalitov
  et~al.}(2015{\natexlab{a}})\citenamefont{Nalitov, Malpuech,
  Ter\ifmmode~\mbox{\c{c}}\else \c{c}\fi{}as, and Solnyshkov}}]{Nalitov2015b}
\bibinfo{author}{\bibfnamefont{A.~V.} \bibnamefont{Nalitov}},
  \bibinfo{author}{\bibfnamefont{G.}~\bibnamefont{Malpuech}},
  \bibinfo{author}{\bibfnamefont{H.}~\bibnamefont{Ter\ifmmode~\mbox{\c{c}}\else
  \c{c}\fi{}as}}, \bibnamefont{and} \bibinfo{author}{\bibfnamefont{D.~D.}
  \bibnamefont{Solnyshkov}}, \bibinfo{journal}{Phys. Rev. Lett.}
  \textbf{\bibinfo{volume}{114}}, \bibinfo{pages}{026803}
  (\bibinfo{year}{2015}{\natexlab{a}}),
  \urlprefix\url{https://link.aps.org/doi/10.1103/PhysRevLett.114.026803}.

\bibitem[{\citenamefont{Nalitov
  et~al.}(2015{\natexlab{b}})\citenamefont{Nalitov, Solnyshkov, and
  Malpuech}}]{Nalitov2015}
\bibinfo{author}{\bibfnamefont{A.~V.} \bibnamefont{Nalitov}},
  \bibinfo{author}{\bibfnamefont{D.~D.} \bibnamefont{Solnyshkov}},
  \bibnamefont{and} \bibinfo{author}{\bibfnamefont{G.}~\bibnamefont{Malpuech}},
  \bibinfo{journal}{Phys. Rev. Lett.} \textbf{\bibinfo{volume}{114}},
  \bibinfo{pages}{116401} (\bibinfo{year}{2015}{\natexlab{b}}),
  \urlprefix\url{https://link.aps.org/doi/10.1103/PhysRevLett.114.116401}.

\bibitem[{\citenamefont{Wu et~al.}(2017)\citenamefont{Wu, Lovorn, and
  MacDonald}}]{MacDonald2017}
\bibinfo{author}{\bibfnamefont{F.}~\bibnamefont{Wu}},
  \bibinfo{author}{\bibfnamefont{T.}~\bibnamefont{Lovorn}}, \bibnamefont{and}
  \bibinfo{author}{\bibfnamefont{A.~H.} \bibnamefont{MacDonald}},
  \bibinfo{journal}{Phys. Rev. Lett.} \textbf{\bibinfo{volume}{118}},
  \bibinfo{pages}{147401} (\bibinfo{year}{2017}),
  \urlprefix\url{https://link.aps.org/doi/10.1103/PhysRevLett.118.147401}.

\bibitem[{\citenamefont{Sala et~al.}(2015)\citenamefont{Sala, Solnyshkov,
  Carusotto, Jacqmin, Lema\^{\i}tre, Ter\ifmmode~\mbox{\c{c}}\else
  \c{c}\fi{}as, Nalitov, Abbarchi, Galopin, Sagnes et~al.}}]{Sala2015}
\bibinfo{author}{\bibfnamefont{V.~G.} \bibnamefont{Sala}},
  \bibinfo{author}{\bibfnamefont{D.~D.} \bibnamefont{Solnyshkov}},
  \bibinfo{author}{\bibfnamefont{I.}~\bibnamefont{Carusotto}},
  \bibinfo{author}{\bibfnamefont{T.}~\bibnamefont{Jacqmin}},
  \bibinfo{author}{\bibfnamefont{A.}~\bibnamefont{Lema\^{\i}tre}},
  \bibinfo{author}{\bibfnamefont{H.}~\bibnamefont{Ter\ifmmode~\mbox{\c{c}}\else
  \c{c}\fi{}as}}, \bibinfo{author}{\bibfnamefont{A.}~\bibnamefont{Nalitov}},
  \bibinfo{author}{\bibfnamefont{M.}~\bibnamefont{Abbarchi}},
  \bibinfo{author}{\bibfnamefont{E.}~\bibnamefont{Galopin}},
  \bibinfo{author}{\bibfnamefont{I.}~\bibnamefont{Sagnes}},
  \bibnamefont{et~al.}, \bibinfo{journal}{Phys. Rev. X}
  \textbf{\bibinfo{volume}{5}}, \bibinfo{pages}{011034} (\bibinfo{year}{2015}),
  \urlprefix\url{https://link.aps.org/doi/10.1103/PhysRevX.5.011034}.

\bibitem[{\citenamefont{Bleu}(2018)}]{thesis2018bleu}
\bibinfo{author}{\bibfnamefont{O.}~\bibnamefont{Bleu}}, Ph.D. thesis,
  \bibinfo{school}{Université Clermont Auvergne} (\bibinfo{year}{2018}),
  \urlprefix\url{http://www.theses.fr/2018CLFAC044}.

\bibitem[{\citenamefont{McCann and Koshino}(2013)}]{mccann2013electronic}
\bibinfo{author}{\bibfnamefont{E.}~\bibnamefont{McCann}} \bibnamefont{and}
  \bibinfo{author}{\bibfnamefont{M.}~\bibnamefont{Koshino}},
  \bibinfo{journal}{Reports on Progress in physics}
  \textbf{\bibinfo{volume}{76}}, \bibinfo{pages}{056503}
  (\bibinfo{year}{2013}).

\bibitem[{\citenamefont{Lopes~dos Santos et~al.}(2007)\citenamefont{Lopes~dos
  Santos, Peres, and Castro~Neto}}]{dosSantos2007}
\bibinfo{author}{\bibfnamefont{J.~M.~B.} \bibnamefont{Lopes~dos Santos}},
  \bibinfo{author}{\bibfnamefont{N.~M.~R.} \bibnamefont{Peres}},
  \bibnamefont{and} \bibinfo{author}{\bibfnamefont{A.~H.}
  \bibnamefont{Castro~Neto}}, \bibinfo{journal}{Phys. Rev. Lett.}
  \textbf{\bibinfo{volume}{99}}, \bibinfo{pages}{256802}
  (\bibinfo{year}{2007}),
  \urlprefix\url{https://link.aps.org/doi/10.1103/PhysRevLett.99.256802}.

\bibitem[{\citenamefont{Moon and Koshino}(2012)}]{Moon2012}
\bibinfo{author}{\bibfnamefont{P.}~\bibnamefont{Moon}} \bibnamefont{and}
  \bibinfo{author}{\bibfnamefont{M.}~\bibnamefont{Koshino}},
  \bibinfo{journal}{Phys. Rev. B} \textbf{\bibinfo{volume}{85}},
  \bibinfo{pages}{195458} (\bibinfo{year}{2012}),
  \urlprefix\url{https://link.aps.org/doi/10.1103/PhysRevB.85.195458}.

\bibitem[{\citenamefont{Bistritzer and MacDonald}(2011)}]{bistritzer2011moire}
\bibinfo{author}{\bibfnamefont{R.}~\bibnamefont{Bistritzer}} \bibnamefont{and}
  \bibinfo{author}{\bibfnamefont{A.~H.} \bibnamefont{MacDonald}},
  \bibinfo{journal}{Proceedings of the National Academy of Sciences}
  \textbf{\bibinfo{volume}{108}}, \bibinfo{pages}{12233}
  (\bibinfo{year}{2011}).

\bibitem[{\citenamefont{Dresselhaus}(1974)}]{Dresselhaus1974}
\bibinfo{author}{\bibfnamefont{G.}~\bibnamefont{Dresselhaus}},
  \bibinfo{journal}{Phys. Rev. B} \textbf{\bibinfo{volume}{10}},
  \bibinfo{pages}{3602} (\bibinfo{year}{1974}),
  \urlprefix\url{https://link.aps.org/doi/10.1103/PhysRevB.10.3602}.

\bibitem[{\citenamefont{Tarnopolsky et~al.}(2019)\citenamefont{Tarnopolsky,
  Kruchkov, and Vishwanath}}]{Tarnopolsky2019}
\bibinfo{author}{\bibfnamefont{G.}~\bibnamefont{Tarnopolsky}},
  \bibinfo{author}{\bibfnamefont{A.~J.} \bibnamefont{Kruchkov}},
  \bibnamefont{and}
  \bibinfo{author}{\bibfnamefont{A.}~\bibnamefont{Vishwanath}},
  \bibinfo{journal}{Phys. Rev. Lett.} \textbf{\bibinfo{volume}{122}},
  \bibinfo{pages}{106405} (\bibinfo{year}{2019}),
  \urlprefix\url{https://link.aps.org/doi/10.1103/PhysRevLett.122.106405}.

\bibitem[{sup()}]{suppl}
\bibinfo{note}{See Supplemental Material at [URL will be inserted by
  publisher].}

\bibitem[{\citenamefont{Bleu et~al.}(2017)\citenamefont{Bleu, Solnyshkov, and
  Malpuech}}]{Bleu2017x}
\bibinfo{author}{\bibfnamefont{O.}~\bibnamefont{Bleu}},
  \bibinfo{author}{\bibfnamefont{D.~D.} \bibnamefont{Solnyshkov}},
  \bibnamefont{and} \bibinfo{author}{\bibfnamefont{G.}~\bibnamefont{Malpuech}},
  \bibinfo{journal}{Phys. Rev. B} \textbf{\bibinfo{volume}{95}},
  \bibinfo{pages}{115415} (\bibinfo{year}{2017}),
  \urlprefix\url{https://link.aps.org/doi/10.1103/PhysRevB.95.115415}.

\bibitem[{\citenamefont{Hatsugai}(1993)}]{Hatsugai1993}
\bibinfo{author}{\bibfnamefont{Y.}~\bibnamefont{Hatsugai}},
  \bibinfo{journal}{Phys. Rev. Lett.} \textbf{\bibinfo{volume}{71}},
  \bibinfo{pages}{3697} (\bibinfo{year}{1993}),
  \urlprefix\url{https://link.aps.org/doi/10.1103/PhysRevLett.71.3697}.

\end{thebibliography}

\renewcommand{\thefigure}{S\arabic{figure}}
\renewcommand{\theequation}{S\arabic{equation}}

\section{Supplementary: Topological moiré polaritons}

In this Supplementary, we present the results of additional calculations. We analyze the symmetry of the moiré unit cell. We show the behavior of the size of the minigaps on the TE-TM splitting. We also present the calculated Chern numbers of all bands for one particular configuration.

\begin{figure}[tbp]
\centering
\includegraphics[width=0.99\linewidth]{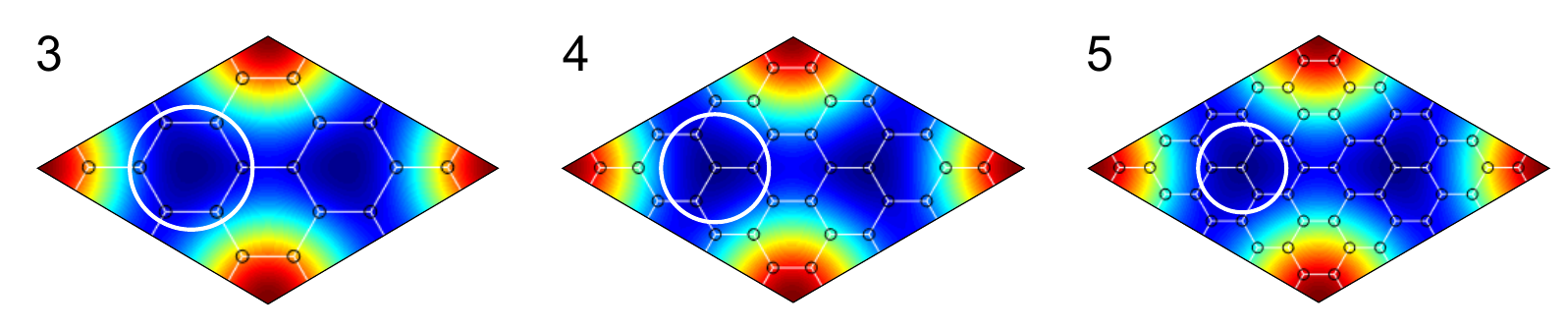}
\caption{Moiré unit cell for $m=3,4,5$ as indicated in panels. The white circles mark the high-symmetry points, whose local symmetry is periodically modified, alternating between hexagonal and two alternative trigonal orientations.
\label{figSrsp}}
\end{figure}

In the main text, we have demonstrated a periodic behavior of the band width accompained with periodically alternating distribution of the Berry curvature. Both effects take their origin in the local symmetry of the  microscopic structure of the moiré unit cell, as shown in Fig.~\ref{figSrsp}. The three panels of this figure correspond to the same values of the moiré parameter $m=3,4,5$, as the panels of Fig.~2(b) of the main text that show the Berry curvature. Fig.~\ref{figSrsp} shows the modulated TE-TM splitting as a false color, together with the positions of the atoms and the links between them. The high-symmetry points (the centers of the two equilateral triangles forming the unit cell) are surrounded by white circles, allowing to track the alternating modification of the local symmetry of the atomic structure. For $m=3$, the local symmetry is hexagonal ($C_6$), for $m=4$ it is trigonal ($C_3$), and for $m=5$ it is also trigonal ($C_3$), but the orientation of the atomic triangle is inverted with respect to $m=4$. This modification of the local symmetry determines the Berry curvature distribution and the width of the band.

\begin{figure}[tbp]
\centering
\includegraphics[width=0.99\linewidth]{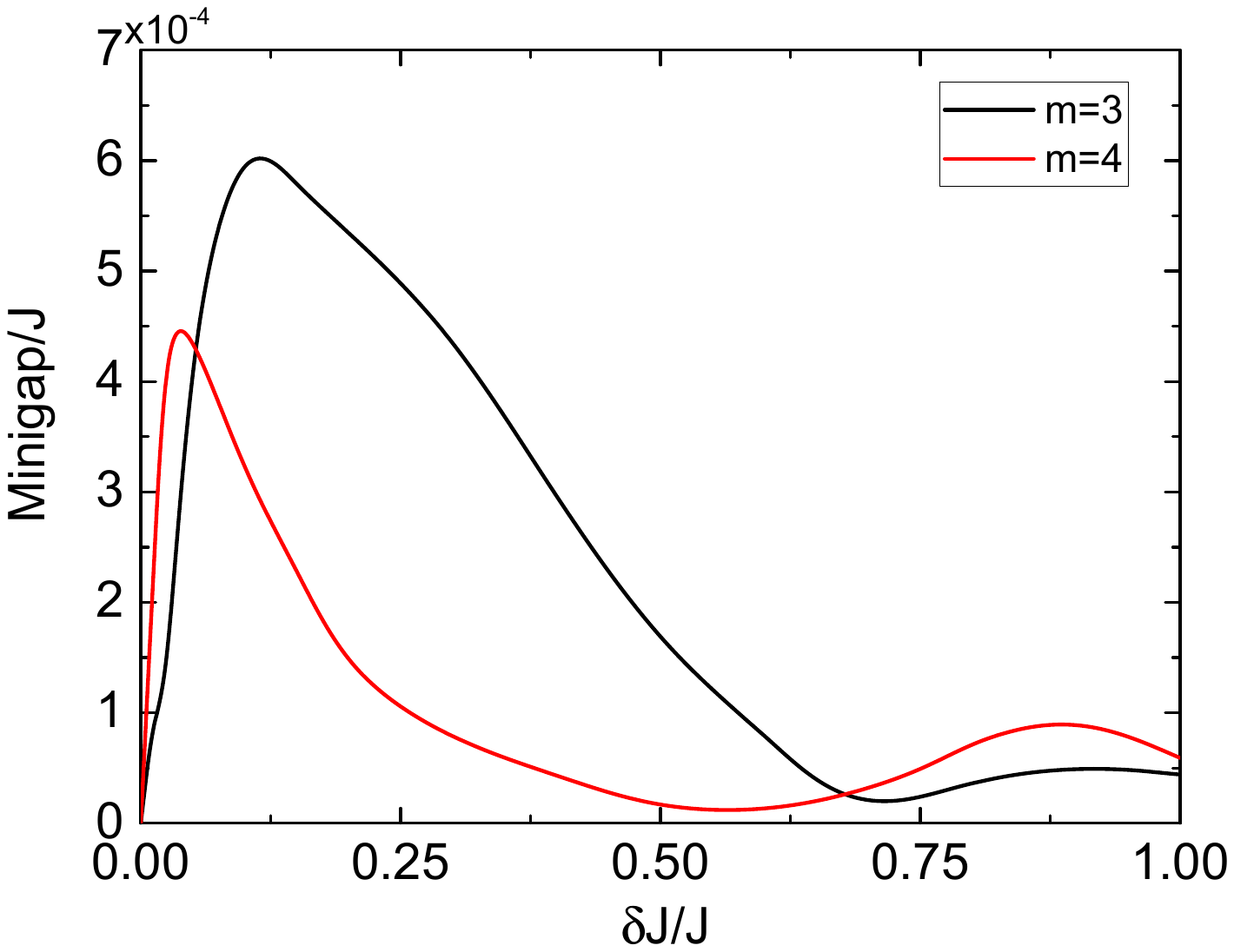}
\caption{Size of the minigap between the 1st and the 2nd bands as counted from the central gap for $m=3$ (black) and $m=4$ (red) as a function of the TE-TM splitting strength $\delta J$ (other parameters as in the main text).
\label{figS1}}
\end{figure}

As stated in the main text, when the moiré parameter $m$ is increased, the size of all bands is reduced as well. The minigaps between these bands are even smaller in magnitude than the bands. Even though these gaps are not necessarily full ones, which means that the allowed bands actually overlap and form a continuum, the bands are nevertheless separated from each other: there are no degenerate eigenvalues at any $k$ if both Zeeman splitting $\Delta$ and the TE-TM splitting $\delta J$ are non-zero. This allows to calculate the Chern numbers of all bands, even though this task is numerically extremely demanding because of the rapid $m^2$ growth of the number of bands and of the required high precision.

In Fig.~\ref{figS1} we show the size of the minigaps as a function of one of the parameters controlling them, the TE-TM splitting $\delta J$ for two smallest moiré parameters $m=3,4$ (black, red). We choose one particular minigap, separating the 1st and the 2nd bands as counted from the central gap.
Comparing the scale of the values of the minigap with the sizes of the bands shown in Fig.~2 of the main text, we see that the latter are much larger. The minigaps are also completely invisible in Fig.~1(e), while the bands exhibit a relatively large width. This is why it is important to check that these minigaps really exist, and that the bands are not crossing. The values shown in Fig.~\ref{figS1} remain well above the numerical uncertainty estimated as $10^{-7}$.

\begin{figure}[tbp]
\centering
\includegraphics[width=0.99\linewidth]{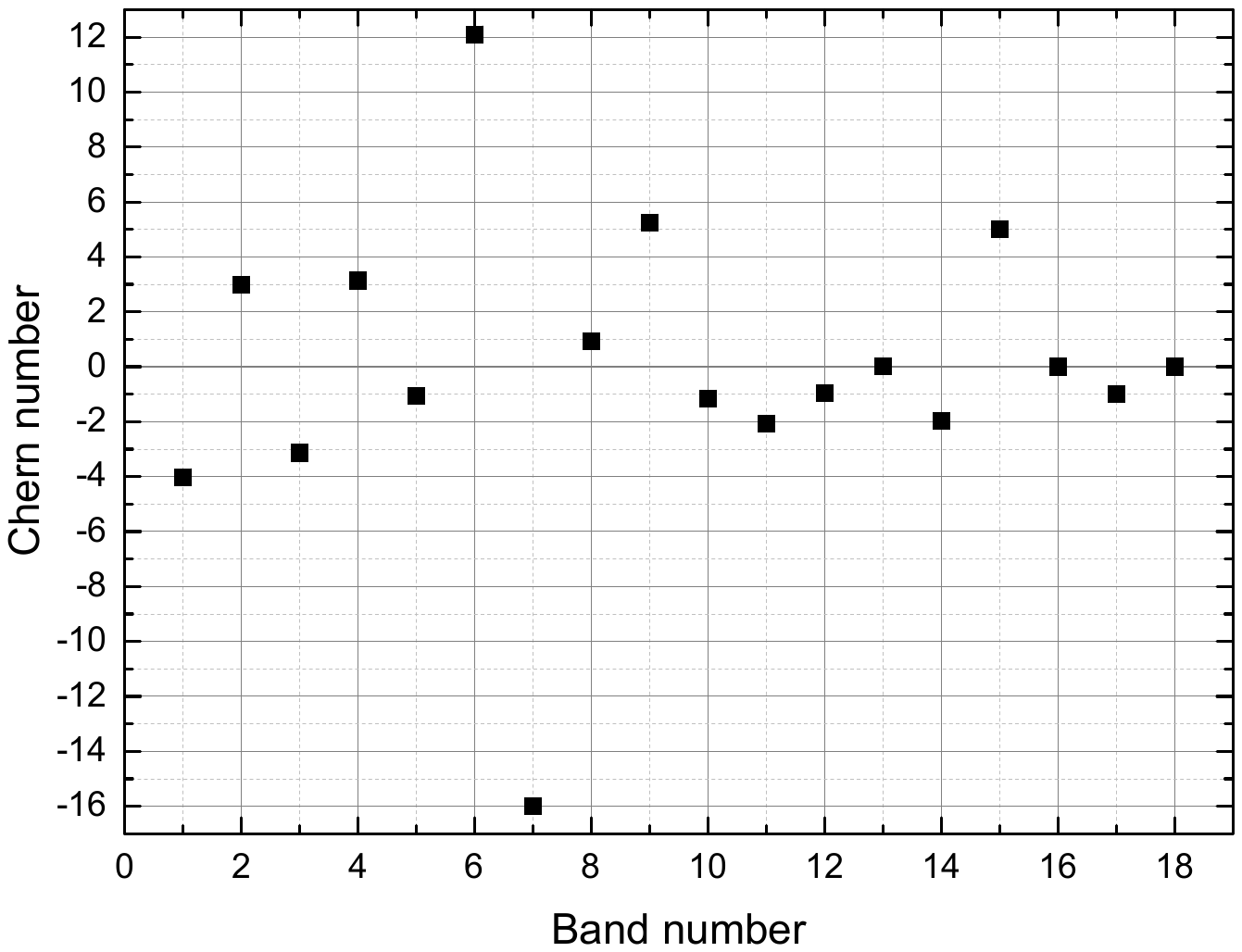}
\caption{Chern numbers for all 18 bands above the central bandgap, in the case of $m=3$ (other parameters as in the main text). The sum of these Chern numbers $|\sum\mathcal{C}_i|=2$.
\label{figS2}}
\end{figure}

Figure~\ref{figS2} shows the calculated Chern numbers for all bands in the case of $m=3$. The Berry curvature of the 1st band in this case is presented in Fig.~2(b3) of the main text. Almost all bands are topologically non-trivial, and some of them exhibit very large Chern numbers (up to $|\mathcal{C}|=16$). We stress that the Chern number of the central bandgap equal to the sum of these Chern numbers gives $|\sum\mathcal{C}_i|=2$, as expected from the topological properties of the two ingredients responsible for the non-trivial behavior: the TE-TM splitting and the Zeeman splitting.

\end{document}